\documentclass[conference]{IEEEtran}
\usepackage{graphicx}
\usepackage{amsmath}
\usepackage{array}
\newcommand{\PreserveBackslash}[1]{\let\temp=\\#1\let\\=\temp}
\newcolumntype{C}[1]{>{\PreserveBackslash\centering}p{#1}}
\newcolumntype{R}[1]{>{\PreserveBackslash\raggedleft}p{#1}}
\newcolumntype{L}[1]{>{\PreserveBackslash\raggedright}p{#1}}
\usepackage[usenames]{color}
\usepackage{colortbl,booktabs}
\usepackage{tabularx}
\usepackage{multicol}
\usepackage{booktabs}
\usepackage[Symbol]{upgreek}
\usepackage{subfigure}
\usepackage{bm}
\usepackage{cite}
\usepackage{balance}
\usepackage[boxed]{algorithm2e}

\usepackage{threeparttable}
\usepackage{amsmath}
\usepackage{dcolumn}
\usepackage{multirow}
\usepackage{booktabs}
\usepackage{amsfonts}
\newcolumntype{d}[1]{D{.}{.}{#1}}

\def \qed {\hfill \vrule height6pt width 6pt depth 0pt}

\begin{document}

\bibliographystyle{IEEEtran} 
\title{Beamspace Channel Estimation for Millimeter-Wave Massive MIMO Systems with Lens Antenna Array}

\author{\IEEEauthorblockN{Linglong Dai$^\ast$, Xinyu Gao$^\ast$, Shuangfeng Han$^\dagger$, Chih-Lin I$^\dagger$, and Xiaodong Wang${^\ddag}$}
\IEEEauthorblockA{$^\ast$Tsinghua National Laboratory for Information Science and Technology (TNList), \\Department of Electronic Engineering, Tsinghua University, Beijing 100084, China\\
$^\dagger$Green Communication Research Center, China Mobile Research
Institute, Beijing 100053, China\\
${^\ddag}$Department of Electrical Engineering, Columbia University, New York, NY 10027, USA\\
Email: daill@tsinghua.edu.cn
}}



\maketitle

\begin{abstract}
By employing the lens antenna array, beamspace MIMO can utilize beam selection to reduce the number of required RF chains in mmWave massive MIMO systems without obvious performance loss. However, to achieve the capacity-approaching performance, beam selection requires the accurate information of beamspace channel of large size, which is challenging, especially when the number of RF chains is limited. To solve this problem, in this paper we propose a reliable support detection (SD)-based channel estimation scheme. Specifically, we propose to decompose the total beamspace channel estimation problem into a series of sub-problems, each of which only considers one sparse channel component. For each channel component, we first reliably detect its support by utilizing the structural characteristics of mmWave beamspace channel. Then, the influence of this channel component is removed from the total beamspace channel estimation problem. After the supports of all channel components have been detected, the nonzero elements of the sparse beamspace channel can be estimated with low pilot overhead. Simulation results show that the proposed SD-based channel estimation outperforms conventional schemes and enjoys satisfying accuracy, even in the low SNR region.
\end{abstract}

\section{Introduction}\label{S1}

\IEEEPARstart The integration of millimeter-wave (mmWave) and massive multiple-input multiple-output (MIMO) has been considered as a key technique for future 5G wireless communications~\cite{han2015large}, since it can achieve significant increase in data rates due to its wider bandwidth and higher spectral efficiency.

However, realizing mmWave massive MIMO in practice is not a trivial task. One key challenging problem is that each antenna in MIMO systems usually requires one dedicated radio-frequency (RF) chain (including digital-to-analog converter, up converter, etc.). This results in unaffordable hardware complexity and energy consumption in mmWave massive MIMO systems, as the number of antennas becomes huge~\cite{han2015large} and the energy consumption of RF chain is high at mmWave frequencies~\cite{alkhateeb2014mimo}. To reduce the  number of required RF chains, the concept of beamspace MIMO has been recently proposed in the pioneering work~\cite{brady2013beamspace}. By employing the lens antenna array instead of the conventional electromagnetic antenna array, beamspace MIMO can transform the conventional spatial channel into beamspace channel by concentrating the signals from different directions (beams) on different antennas~\cite{brady2013beamspace}. Since the scattering in mmWave communications is not rich, the number of effective prorogation paths is quite limited~\cite{han2015large}, occupying only a small number of beams. As a result, the mmWave beamspace channel is sparse~\cite{brady2013beamspace}, and we can select a small number of dominant beams to significantly reduce the dimension of MIMO system and the number of required RF chains without obvious performance loss~\cite{sayeed2013beamspace}.

Nevertheless, beam selection requires the base station (BS) to acquire the information of beamspace channel of large size, which is challenging, especially when the number of RF chains is limited. To solve this problem, some advanced schemes based on compressive sensing (CS) have been proposed very recently~\cite{alkhateeb2014channel,alkhateeb2015compressed,love15}. The key idea of these schemes is to utilize the sparsity of mmWave channels in the angle domain to efficiently estimate the mmWave massive MIMO channel of large size. However, these schemes are designed for hybrid precoding systems~\cite{gao15energy}, where the phase shifter network can generate beams with sufficiently high angle resolution to improve the channel estimation accuracy. By contrast, in beamspace MIMO systems, although the phase shifter network can be replaced by lens antenna array to further reduce the hardware cost and energy consumption, the generated beams are predefined with a fixed yet limited angle resolution. If we directly apply the existing channel estimation schemes to beamspace MIMO systems with lens antenna array, the performance will be not very satisfying~\cite{brady2013beamspace}. To the best of our knowledge, the channel estimation problem for beamspace MIMO systems has not been well addressed in the literature.

In this paper, by fully utilizing the structural characteristics of mmWave beamspace channel, we propose a reliable support detection (SD)-based channel estimation scheme. The basic idea is to decompose the total beamspace channel estimation problem into a series of sub-problems, each of which only considers one sparse channel component (a vector containing the information of a specific propagation direction). For each channel component, we first detect its support (i.e., the index set of nonzero elements in a sparse vector) according to the estimated position of the strongest element. Then, the influence of this channel component is removed from the total beamspace channel estimation problem, and the support of the next channel component  is detected in a similar method. After the supports of all channel components have been detected, the nonzero elements of the sparse beamspace channel can be estimated with low pilot overhead. Simulation results show that the proposed SD-based channel estimation outperforms conventional schemes, especially in the low signal-to-noise ratio (SNR) region, which is more attractive for  mmWave massive MIMO systems where low SNR is the typical case before beamforming~\cite{alkhateeb2014channel}.

{\it Notation}: Lower-case and upper-case boldface letters denote vectors and matrices, respectively; ${( \cdot )^H}$, ${( \cdot )^{ - 1}}$,  and ${{\rm{tr}}( \cdot )}$ denote the conjugate transpose, inversion, and trace of a matrix, respectively; ${{\left\|  \cdot  \right\|_F}}$  denotes the Frobenius norm of a matrix; ${\left|  \cdot  \right|}$ denotes the amplitude of a scalar; ${{\rm{Card}}\left(  \cdot  \right)}$  denotes the cardinality of a set; Finally, ${{{\mathbf{I}}_{K}}}$ is the ${K \times K}$ identity matrix.


\section{System Model}\label{S2}
In this paper, we consider a typical mmWave massive MIMO system working in time division duplexing (TDD) model, where the BS employs ${N}$  antennas and ${{N_{{\rm{RF}}}}}$ RF chains to simultaneously serve ${K}$  single-antenna users~\cite{brady2013beamspace,sayeed2013beamspace}.  As shown in Fig. 1 (a), for conventional MIMO systems in the spatial domain, the  ${K \times 1}$ received signal vector ${{{\mathbf{y}}^{\text{DL}}}}$ for all ${K}$ users in the downlink can be presented by
\begin{equation}\label{eq1}
{{\mathbf{y}}^{\text{DL}}}={{\mathbf{H}}^{H}}\mathbf{Ps}+\mathbf{n},
\end{equation}
where ${{{\mathbf{H}}^{H}}\in {{\mathbb{C}}^{K\times N}}}$ is the downlink channel matrix, ${\mathbf{H}=\left[ {{\mathbf{h}}_{1}},{{\mathbf{h}}_{2}},\cdots,{{\mathbf{h}}_{K}} \right]}$ is the uplink channel matrix according to the channel reciprocity~\cite{love15}, ${{{\mathbf{h}}_{k}}}$ of size ${N \times 1}$ is the channel vector between the BS and the ${k}$th user, ${{\bf{s}}}$ of size ${K \times 1}$ is the signal vector for all ${K}$  users with normalized power ${\mathbb{E}\left( \mathbf{s}{{\mathbf{s}}^{H}} \right)={{\mathbf{I}}_{K}}}$, ${{\bf{P}}}$ of size ${N \times K}$ is the precoding matrix satisfying the total transmit power constraint ${\rho}$ as ${{\rm{tr}}\left( {{\bf{P}}{{\bf{P}}^H}} \right) \le \rho }$. Finally, ${\mathbf{n}\sim\mathcal{C}\mathcal{N}\left( 0,\sigma _{\text{DL}}^{2}{{\mathbf{I}}_{K}} \right)}$ is the ${K \times 1}$ noise vector, where ${{\sigma _{{\rm{DL}}}^2}}$ is the downlink noise power.

\subsection{MmWave channel model}\label{S2.1}
In this paper, we adopt the widely used Saleh-Valenzuela channel model to embody the low rank and spatial correlation characteristics of mmWave communications as~\cite{brady2013beamspace,sayeed2013beamspace,alkhateeb2014channel,alkhateeb2015compressed,love15}
\begin{equation}\label{eq2}
{{\bf{h}}_k} = \sqrt {\frac{N}{{L + 1}}} \sum\limits_{i = 0}^L {\beta _k^{\left( i \right)}{\bf{a}}\left( {\psi _k^{\left( i \right)}} \right)}  = \sqrt {\frac{N}{{L + 1}}} \sum\limits_{i = 0}^L {{{\bf{c}}_i}} ,
\end{equation}
where ${{{\mathbf{c}}_{0}}=\beta _{k}^{\left( 0 \right)}\mathbf{a}\left( \psi _{k}^{\left( 0 \right)} \right)}$ is the line-of-sight (LoS) component of the ${k}$th user with ${\beta _k^{\left( 0 \right)}}$ presenting the complex gain and ${{\psi _k^{\left( 0 \right)}}}$ denoting the spatial direction, ${{{\mathbf{c}}_{i}}=\beta _{k}^{\left( i \right)}\mathbf{a}\left( \psi _{k}^{\left( i \right)} \right)}$ for ${1\le i\le L}$ is the ${i}$th non-line-of-sight (NLoS) component of the ${k}$th user, and ${L}$ is the total number of NLoS components which can be usually obtained by channel measurement~\cite{rappaport2011state}, ${\mathbf{a}\left( \psi  \right)}$ is the ${N \times 1}$ array steering vector. For the typical uniform linear array (ULA) with ${N}$ antennas, we have
\begin{equation}\label{eq3}
{\bf{a}}\left( \psi  \right) = \frac{1}{{\sqrt N }}{\left[ {{e^{ - j2\pi \psi m}}} \right]_{m \in {\cal I}\left( N \right)}},
\end{equation}
where ${{\cal I}\left( N \right) = \left\{ {l - \left( {N - 1} \right)/2,\;l = 0,1, \cdots ,N - 1} \right\}}$ is a symmetric set of indices centered around zero. The spatial direction is defined as ${\psi  \buildrel \Delta \over = \frac{d}{\lambda }\sin \theta }$~\cite{brady2013beamspace}, where ${\theta }$ is the physical direction, ${\lambda }$ is the signal wavelength, and ${d}$ is the antenna spacing which usually satisfies ${d = \lambda /2}$ at mmWave frequencies.


\begin{figure}[tp]
\begin{center}
\vspace*{-3mm}\includegraphics[width=1\linewidth]{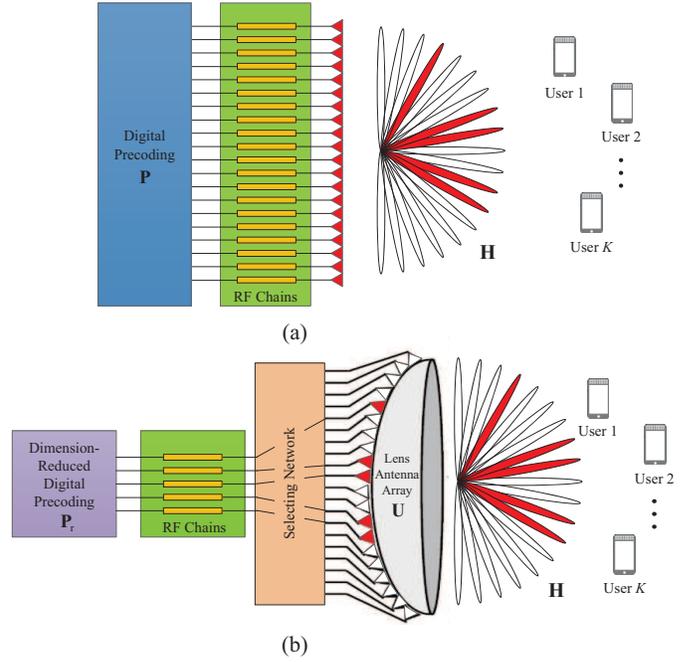}
\end{center}
\vspace*{-3mm}\caption{System architectures: (a) conventional MIMO; (b) beamspace MIMO.} \label{FIG1}
\end{figure}

\subsection{Beamspace  MIMO}\label{S2.2}
The conventional channel (\ref{eq2}) in the spatial domain can be transformed to the beamspace channel by employing a carefully designed lens antenna array~\cite{brady2013beamspace} as shown in Fig. 1 (b). Specifically, such lens antenna array plays the role of a spatial discrete fourier transform (DFT) matrix ${{\bf{U}}}$, which contains the array steering vectors of ${N}$ orthogonal directions (beams) covering the entire angle space as~\cite{brady2013beamspace}
\begin{equation}\label{eq4}
{\bf{U}} = {\left[ {{\bf{a}}\left( {{{\bar \psi }_1}} \right),{\bf{a}}\left( {{{\bar \psi }_2}} \right), \cdots ,{\bf{a}}\left( {{{\bar \psi }_N}} \right)} \right]^H},
\end{equation}
where ${{\bar \psi _n} = \frac{1}{N}\left( {n - \frac{{N + 1}}{2}} \right)}$ for ${n = 1,2, \cdots ,N}$. Then, by assuming ${{N_{{\rm{RF}}}}=K}$ without loss of generality, the system model of beamspace MIMO can be represented by
\begin{equation}\label{eq5}
{{\bf{\tilde y}}^{{\rm{DL}}}} = {{\bf{H}}^H}{{\bf{U}}^H}{\bf{B}}{{\bf{P}}_{\rm{r}}}{\bf{s}} + {\bf{n}} = {{\bf{\tilde H}}^H}{\bf{B}}{{\bf{P}}_{\rm{r}}}{\bf{s}} + {\bf{n}},
\end{equation}
where ${{{\mathbf{\tilde{y}}}^{\text{DL}}}}$ is the received downlink signal vector in the beamspace, ${{{\bf{\tilde H}}^H} = {{\bf{H}}^H}{{\bf{U}}^H} = {\left( {{\bf{UH}}} \right)^H}}$ is defined as the downlink beamspace channel matrix whose ${N}$ columns correspond to ${N}$ orthogonal beams, ${{\bf{B}}}$ of size ${N\times K}$ is the selecting matrix whose entries belong to ${\left\{ {0,1} \right\}}$. For example, if the ${n}$th beam is selected by the ${k}$th user, the element of ${{\bf{B}}}$ at the ${n}$th row and the ${k}$th column is 1.  Finally, ${{{\bf{P}}_{\rm{r}}}}$ of size ${K \times K}$ is the dimension-reduced digital precoding matrix. It is worth pointing out that the beamspace channel ${{{\mathbf{\tilde{H}}}^{H}}}$ (or equivalently ${\mathbf{\tilde{H}}}$) has a sparse structure~\cite{brady2013beamspace,sayeed2013beamspace} due to the limited number of dominant scatters in the mmWave prorogation environments~\cite{han2015large}. Therefore, we can select only a small number of appropriate beams according to the sparse beamspace channel  to significantly reduce the effective channel dimension without obvious performance loss. Consequently, only a small-size digital precoder ${{{\bf{P}}_{\rm{r}}}}$ is required, leading to a small number of required RF chains. Unfortunately, acquiring the beamspace channel of large size in practice is challenging, especially when the number of RF chains is limited.


\section{Beamspace Channel Estimation}\label{S3}
In this section, based on the beamspace MIMO architecture, we first introduce a pilot transmission strategy. After that, an adaptive selecting network is designed to obtain the measurements of the beamspace channel. Finally, a SD-based channel estimation is proposed to estimate the beamspace channel with limited number of RF chains and low pilot overhead.

\subsection{Pilot transmission}\label{S3.1}
To estimate the beamspace channel, in the uplink of TDD systems, all users need to transmit the known pilot sequences to the BS over ${Q}$ instants (each user transmits one pilot symbol in each instant) for channel estimation, and we assume that the beamspace channel remains unchanged within such channel coherence time (i.e., ${Q}$ instants)~\cite{tse2005fundamentals}. In this paper, we consider the pilot transmission strategy, where ${Q}$ instants are divided into ${M}$ blocks and each block consists of ${K}$ instants, i.e., ${Q = MK}$. For the ${m}$th block, we define ${{{\bf{\Psi }}_m}}$ of size ${K \times K}$ as the pilot matrix, which contains ${K}$ mutually orthogonal pilot sequences transmitted by ${K}$ users over ${K}$ instants~\cite{tse2005fundamentals}. Obviously, we have ${{{\bf{\Psi }}_m}{\bf{\Psi }}_m^H = {{\bf{I}}_K}}$ and ${{\bf{\Psi }}_m^H{{\bf{\Psi }}_m} = {{\bf{I}}_K}}$.

Then, according to Fig. 1 (b) and the channel reciprocity~\cite{love15} in TDD systems, the received uplink signal matrix ${\mathbf{\tilde{Y}}_{m}^{\text{UL}}}$ of size ${N \times K}$ at the BS in the ${m}$th block can be presented as
\begin{equation}\label{eq8}
\mathbf{\tilde{Y}}_{m}^{\text{UL}}\!=\!\mathbf{UH}{{\mathbf{\Psi }}_{m}}\!+\!{{\mathbf{N}}_{m}}\!=\!\mathbf{\tilde{H}}{{\mathbf{\Psi }}_{m}}\!+\!{{\mathbf{N}}_{m}},\quad m\!=\!1,2,\cdots,M,
\end{equation}
where ${{{\mathbf{N}}_{m}}}$ is the ${N \times K}$ noise matrix in the ${m}$th block, whose entries are independent and identically distributed (i.i.d.) complex Gaussian random variables with mean zero and variance ${{\sigma _{{\rm{UL}}}^2}}$ (the uplink noise power).

\subsection{Measurements of the beamspace channel}\label{S3.2}
We consider the ${m}$th block without loss of generality. During the pilot transmission, the BS should employ a combiner ${{\bf{W}}_m}$ of size ${K \times N}$ to combine the received uplink signal matrix ${\mathbf{\tilde{Y}}_{m}^{\text{UL}}}$~(\ref{eq8}). Then, we can obtain ${{{\bf{R}}_m}}$ of size ${K \times K}$ in the baseband sampled by ${{N_{{\rm{RF}}}} = K}$ RF chains as
\begin{equation}\label{eq13}
{{\mathbf{R}}_{m}}={{\mathbf{W}}_{m}}\mathbf{\tilde{Y}}_{m}^{\text{UL}}={{\mathbf{W}}_{m}}\mathbf{\tilde{H}}{{\mathbf{\Psi }}_{m}}+{{\mathbf{W}}_{m}}{{\mathbf{N}}_{m}}.
\end{equation}
After that, by multiplying the known pilot matrix ${{\bf{\Psi }}_m^H}$ on the right side of~(\ref{eq13}), the ${K \times K}$ measurement matrix ${{{\bf{Z}}_m}}$ of the beamspace channel ${\mathbf{\tilde{H}}}$ can be obtained by
\begin{equation}\label{eq14}
{{\bf{Z}}_m} = {{\bf{R}}_m}{\bf{\Psi }}_m^H = {{\bf{W}}_m}{\bf{\tilde H}} + {\bf{N}}_m^{{\rm{eff}}},
\end{equation}
where ${\mathbf{N}_{m}^{\text{eff}}={{\mathbf{W}}_{m}}{{\mathbf{N}}_{m}}\mathbf{\Psi }_{m}^{H}}$ is the effective noise matrix.

Note that here we focus on estimating the beamspace channel ${{{\bf{\tilde h}}_k}}$ of the ${k}$th user without loss of generality, and the similar method can be directly applied to other users to obtain the complete beamspace channel ${{{\mathbf{\tilde{H}}}}}$. Then, after ${M}$ blocks for the pilot transmission, we can obtain an ${Q \times 1}$ measurement vector ${{{\bf{\bar z}}_k}}$ for ${{{\bf{\tilde h}}_k}}$ as
\begin{equation}\label{eq15}
{{\bf{\bar z}}_k} = \left[ {\begin{array}{*{20}{c}}
{{{\bf{z}}_{1,k}}}\\
{{{\bf{z}}_{2,k}}}\\
 \vdots \\
{{{\bf{z}}_{M,k}}}
\end{array}} \right] = \left[ {\begin{array}{*{20}{c}}
{{{\bf{W}}_1}}\\
{{{\bf{W}}_2}}\\
 \vdots \\
{{{\bf{W}}_M}}
\end{array}} \right]  {{\bf{\tilde h}}_k} + \left[ {\begin{array}{*{20}{c}}
{{\bf{n}}_{1,k}^{{\rm{eff}}}}\\
{{\bf{n}}_{2,k}^{{\rm{eff}}}}\\
 \vdots \\
{{\bf{n}}_{M,k}^{{\rm{eff}}}}
\end{array}} \right] \buildrel \Delta \over = {\bf{\bar W}}{{\bf{\tilde h}}_k} + {{\bf{\bar n}}_k},
\end{equation}
where ${{{\bf{z}}_{m,k}}}$, ${{{\bf{\tilde h}}_k}}$, and ${{\bf{n}}_{m,k}^{{\rm{eff}}}}$ are the ${k}$th column of ${{{\bf{Z}}_m}}$, ${{\bf{\tilde H}}}$, and ${{\bf{N}}_m^{{\rm{eff}}}}$ in~(\ref{eq14}), respectively. ${{{{\bf{\bar z}}}_k}}$, ${{{\bf{\bar W}}}}$, and ${{{{\bf{\bar n}}}_k}}$ are of size ${Q \times 1}$, ${Q \times N}$, and ${Q \times 1}$, respectively. Our target is to reliably reconstruct ${{{\bf{\tilde h}}_k}}$ based on ${{{\bf{\bar z}}_k}}$ with the pilot overhead ${Q}$ as low as possible. However, if we directly utilize the  traditional selecting network in beamspace MIMO systems as shown in Fig. 1 (b) to design ${{{\bf{\bar W}}}}$ (or equivalently ${{\bf{W}}_m}$ for ${m = 1,2, \cdots ,M}$), each row of ${{{\bf{\bar W}}}}$ will have one and only one nonzero element~\cite{amadorilow}. Consequently, to guarantee that the measurement vector ${{{\bf{\bar z}}_k}}$ contains the complete information of the beamspace channel ${{{\bf{\tilde h}}_k}}$, the pilot overhead ${Q}$ should be at least larger than ${N}$, which is unaffordable since the number of antennas ${N}$ is usually huge in mmWave massive MIMO systems as mentioned above.

\begin{figure}[tp]
\begin{center}
\vspace*{-3mm}\includegraphics[width=1\linewidth]{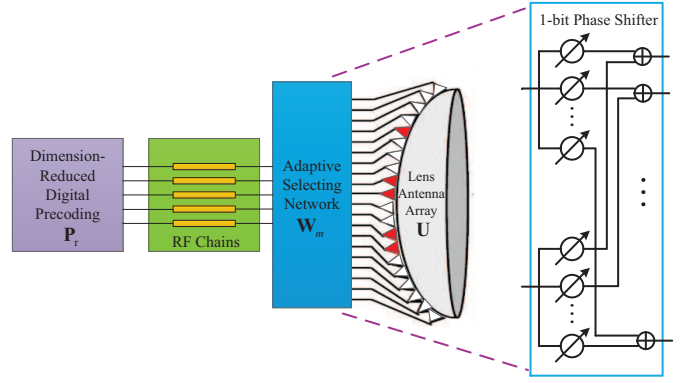}
\end{center}
\vspace*{-3mm}\caption{Proposed adaptive selecting network in beamspace MIMO systems.} \label{FIG1}
\end{figure}

To this end, we propose an adaptive selecting network for mmWave massive MIMO systems with lens antenna array as shown in Fig. 2, where the selecting network in Fig. 1 (b) is replaced by a phase shifter network. During the data transmission, the proposed adaptive selecting network can be configured to realize the traditional function of beam selection\footnote{Specifically, we can turn off some phase shifters to realize ``unselect" and set some phase shifters to shift the phase 0 degree to realize ``select" for beam selection.}. Furthermore, during the beamspace channel estimation, this adaptive selecting network can be also adaptively used as an analog combiner ${{\bf{W}}_m}$~\cite{gao15energy} to combine the uplink signals. With the help of the proposed adaptive selecting network, we can guarantee that ${{{\bf{\bar z}}_k}}$~(\ref{eq15}) has the complete information of ${{{\bf{\tilde h}}_k}}$, even if ${Q < N}$. Moreover, due to the limited number of dominant scatters in the mmWave prorogation environments~\cite{sayeed2013beamspace}, ${{{\bf{\tilde h}}_k}}$ is a sparse vector. Therefore,~(\ref{eq15}) can be formulated as a typical sparse signal recovery problem~\cite{bajwa2010compressed}.

Our next target is to design the analog combiner ${{{\bf{\bar W}}}}$. Under the framework of CS, to achieve the satisfying recovery accuracy, ${{{\bf{\bar W}}}}$ should be designed to make the mutual coherence ${\mu  \buildrel \Delta \over = \mathop {\max }\limits_{i \ne j} \left| {{\bf{\bar w}}_i^H{{{\bf{\bar w}}}_j}} \right|}$ as small as possible, where ${{{{{\bf{\bar w}}}_i}}}$ is the ${i}$th column of ${{{\bf{\bar W}}}}$. There are already some matrices that have been proved to enjoy small ${\mu}$, such as the i.i.d. Gaussian random matrix and Bernoulli random matrix~\cite{bajwa2010compressed}. In our paper, we select the Bernoulli random matrix as the combiner ${{{\bf{\bar W}}}}$, i.e., each element of ${{{\bf{\bar W}}}}$ is randomly selected from ${\frac{1}{\sqrt{Q}}\left\{ -1,+1 \right\}}$ with equal probability. This is due to the facts that: i) all elements of ${{{\bf{\bar W}}}}$ share the same normalized amplitude, which can be realized by phase shifters; ii) the resolution of phase shifter can be only 1 bit, since we only need to shift the phases 0 degree and ${\pi}$ degree. This means that the cost and energy consumption of the phase shifter network can be significantly reduced~\cite{balanis2012antenna}.

\subsection{SD-based channel estimation}\label{S3.1}
After ${{{\bf{\bar W}}}}$ has been designed by the proposed adaptive selecting network,~(\ref{eq15}) can be solved by classical CS algorithms, such as orthogonal matching pursuit (OMP) and compressive sampling matching pursuit (CoSaMP)~\cite{tropp2007signal}. However, when the uplink SNR is low, which is the typical case in mmWave  massive MIMO systems due to the lack of beamforming gain and the low transmit power of users~\cite{alkhateeb2014channel}, ${{{\mathbf{\tilde{h}}}_{k}}}$ will be overwhelmed by noise. As a result, the support of ${{{\mathbf{\tilde{h}}}_{k}}}$ detected by classical CS algorithms is usually inaccurate, leading to the deteriorated performance. In this paper, by utilizing the structural characteristics of mmWave beamspace channel, we propose a SD-based channel estimation, which can detect the support more accurately and achieve better performance than classical CS algorithms, especially in the low SNR region.

In the following \textbf{Lemma 1}, we will first prove a special property of mmWave beamspace channel, which is one of the two bases of the proposed SD-based channel estimation.

\vspace*{+2mm} \noindent\textbf{Lemma 1}. {\it Represent the beamspace channel ${{{\bf{\tilde h}}_k}}$ as ${{{\bf{\tilde h}}_k} = \sqrt {N/\left( {L + 1} \right)} \sum\nolimits_{i = 0}^L {{{{\bf{\tilde c}}}_i}}}$, where ${{{\bf{\tilde c}}_i} = {\bf{U}}{{\bf{c}}_i}}$ is the ${i}$th channel component of ${{{\bf{\tilde h}}_k}}$ in the beamspace. Then, any two channel components ${{{\bf{\tilde c}}_i}}$ and ${{{\bf{\tilde c}}_j}}$ are asymptotically orthogonal when the number of antennas ${N}$  in mmWave massive MIMO systems with lens antenna array tends to infinity, i.e.,}
\begin{equation}\label{eq16}
\mathop {\lim }\limits_{N \to \infty } \left| {{\bf{\tilde c}}_i^H{{{\bf{\tilde c}}}_j}} \right| = 0,\quad \forall \;i,j = 0,1, \cdots ,L,\quad i \ne j.
\end{equation}

\vspace*{+2mm}
\textit{Proof:} Based on~(\ref{eq2})-(\ref{eq4}), we have
\begin{align}\label{eq17}
{\bf{\tilde c}}_i^H{{\bf{\tilde c}}_j} & = \beta _k^{\left( i \right)*}\beta _k^{\left( j \right)}{{\bf{a}}^H}\left( {\psi _k^{\left( i \right)}} \right){{\bf{U}}^H}{\bf{Ua}}\left( {\psi _k^{\left( j \right)}} \right) \\ \nonumber
& = \beta _{k}^{\left( i \right)*}\beta _{k}^{\left( j \right)}\Upsilon \left( \psi _{k}^{\left( i \right)}-\psi _{k}^{\left( j \right)} \right),
\end{align}
where ${\Upsilon \left( x \right) \buildrel \Delta \over = \frac{{\sin N\pi x}}{{N\sin \pi x}}}$. Note that ${{\psi _k^{\left( i \right)}}}$ and ${{\psi _k^{\left( j \right)}}}$ belong to ${\left[ { - 0.5,0.5} \right]}$ according to the definitions in~(\ref{eq3}). Therefore, as long as ${\left( {\psi _k^{\left( i \right)} - \psi _k^{\left( j \right)}} \right) \ne 0\;{\kern 1pt} {\rm{or}}\;{\kern 1pt} 1}$, which can be guaranteed by ${i \ne j}$, we have
\begin{align}\label{eq18}
\left| {\Upsilon \left( {\psi _k^{\left( i \right)} - \psi _k^{\left( j \right)}} \right)} \right| \le \frac{1}{N}\left| {\frac{1}{{\sin \pi \left( {\psi _k^{\left( i \right)} - \psi _k^{\left( j \right)}} \right)}}} \right|.
\end{align}
Based on~(\ref{eq18}), we can conclude that
\begin{align}\label{eq19}
0 \! \le \! \underset{N\to \infty }{\mathop{\lim }}\,\left| {\bf{\tilde c}}_i^H{{\bf{\tilde c}}_j} \right| \! \le \! \underset{N\to \infty }{\mathop{\lim }}\,\frac{1}{N}\left| \frac{\beta _{k}^{\left( i \right)*}\beta _{k}^{\left( j \right)}}{\sin \pi \left( \psi _{k}^{\left( i \right)}-\psi _{k}^{\left( j \right)} \right)} \right| \!= \!0,
\end{align}
which verifies the conclusion~(\ref{eq16}).
\qed

\textbf{Lemma 1} implies that we can decompose the total beamspace channel estimation problem  into a series of independent sub-problems, each of which only considers one specific channel component approximately orthogonal to the others. Specifically, we can first estimate the strongest channel component. After that, we can remove the influence of this component from the total estimation problem, and then the  channel component with the second strongest power can be estimated. Such procedure will be repeated until all ${\left( {L + 1} \right)}$ channel components have been estimated. Next, in the following \textbf{Lemma 2}, we will prove another special structural characteristic of mmWave beamspace channel to show how to estimate each channel component in the beamspace.

\vspace*{+2mm} \noindent\textbf{Lemma 2}. {\it Consider the ${i}$th channel component  ${{{{{\bf{\tilde c}}}_i}}}$ in the beamspace, and assume ${V}$ is an even integer without loss of generality. The ratio between the power ${{{P_V}}}$ of ${V}$ strongest elements of ${{{{{\bf{\tilde c}}}_i}}}$ and the total power ${{{P_T}}}$ of ${{{{{\bf{\tilde c}}}_i}}}$ can be lower-bounded by
\begin{equation}\label{eq20}
\frac{{{P_V}}}{{{P_T}}} \ge \frac{2}{{{N^2}}}\sum\limits_{i = 1}^{V/2} {\frac{1}{{{{\sin }^2}\left( {\frac{{\left( {2i - 1} \right) \pi}}{{2N}}} \right)}}}.
\end{equation}
Moreover, once the position ${n_i^ * }$ of the strongest element of ${{{{{\bf{\tilde c}}}_i}}}$ is determined, the other ${V-1}$ strongest elements will uniformly locate around it with the interval ${1/N}$.}

\begin{figure}[tp]
\begin{center}
\vspace*{-3mm}\includegraphics[width=0.9\linewidth]{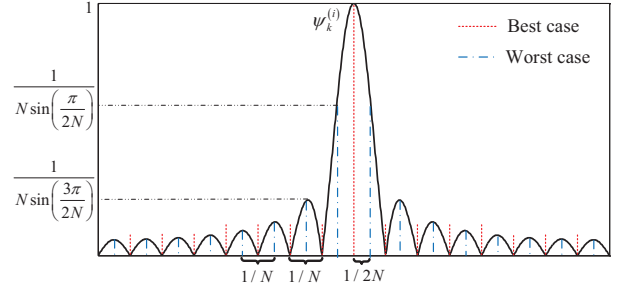}
\end{center}
\vspace*{-4mm}\caption{The normalized amplitude distribution of the elements in ${{{{{\bf{\tilde c}}}_i}}}$.} \label{FIG1}
\end{figure}

\vspace*{+2mm}
\textit{Proof:} Based on~(\ref{eq2})-(\ref{eq4}), the ${i}$th channel component  ${{{{{\bf{\tilde c}}}_i}}}$ in the beamspace can be presented as
\begin{equation}\label{eq21}
{{\mathbf{\tilde{c}}}_{i}}\!=\!\beta _{k}^{\left( i \right)}{{\left[ \Upsilon \left( {{{\bar{\psi }}}_{1}}\!-\!\psi _{k}^{\left( i \right)} \right),\cdots,\Upsilon \left( {{{\bar{\psi }}}_{N}}\!-\!\psi _{k}^{\left( i \right)} \right) \right]}^{H}}.
\end{equation}

Fig. 3 shows the normalized amplitude distribution of the elements in ${{{{{\bf{\tilde c}}}_i}}}$, where the set of red dash lines (or blue dot dash lines) presents the set of spatial directions ${{\bar \psi _n} = \frac{1}{N}\left( {n - \frac{{N + 1}}{2}} \right)}$ for ${n = 1,2, \cdots ,N}$ in~(\ref{eq4}) pre-defined by lens antenna array. From Fig. 3, we can observe that when the practical spatial direction  ${{\psi _k^{\left( i \right)}}}$ exactly equals one pre-defined spatial direction, there is only one strongest element containing all the power of ${{{{{\bf{\tilde c}}}_i}}}$, which is the best case. In contrast, the worst case will happen when the distance between ${{\psi _k^{\left( i \right)}}}$ and one pre-defined spatial direction is equal to ${1/2N}$. In this case, the power ${{{P_V}}}$ of ${V}$ strongest elements of ${{{{{\bf{\tilde c}}}_i}}}$ is
\begin{equation}\label{eq22}
{P_V} = \frac{{2{{\left( {\beta _k^{\left( i \right)}} \right)}^2}}}{{{N^2}}}\sum\limits_{i = 1}^{V/2} {\frac{1}{{{{\sin }^2}\left( {\frac{{\left( {2i - 1} \right)\pi }}{{2N}}} \right)}}}.
\end{equation}

Besides, according to~(\ref{eq21}), the total power ${{P_T}}$ of ${{{{{\bf{\tilde c}}}_i}}}$ can be calculated as ${{P_T} = {\bf{\tilde c}}_i^H{{\bf{\tilde c}}_i} = {\left( {\beta _k^{\left( i \right)}} \right)^2}}$. Therefore, we can conclude that ${{P_V}/{P_T}}$ is lower-bounded by~(\ref{eq20}).
Moreover, as shown in Fig. 3, once the position ${n_i^ * }$ of the strongest element of ${{{{{\bf{\tilde c}}}_i}}}$ is determined, the other ${V-1}$ strongest elements will uniformly locate around it with the interval ${1/N}$. \qed

From \textbf{Lemma 2}, we can derive two important conclusions. The first one is that ${{{{{\bf{\tilde c}}}_i}}}$  can be considered as a sparse vector, since the most power of ${{{{{\bf{\tilde c}}}_i}}}$  is focused on a small number of dominant elements. For example, when ${N=256}$ and ${V=8}$, the lower-bound of ${{P_V}/{P_T}}$ is about 95\%. This means that we can retain only a small number (e.g., ${V=8}$) of elements of ${{{{{\bf{\tilde c}}}_i}}}$ with strong power and regard other elements as zero without obvious performance loss. The second one is that the support of sparse vector ${{{{{\bf{\tilde c}}}_i}}}$ can be uniquely determined by ${n_i^ * }$ as\footnote{Correspondingly, when ${V}$ is odd, the support of ${{{{{\bf{\tilde c}}}_i}}}$ should be ${{\rm{supp}}\left( {{{{\bf{\tilde c}}}_i}} \right) = \,\bmod \,{{\mkern 1mu} _N}\left\{ {n_i^ *  - \frac{{V - 1}}{2}, \cdots ,n_i^ *  + \frac{{V - 1}}{2}} \right\}}$.}
\begin{equation}\label{eq24}
{\rm{supp}}\left( {{{{\bf{\tilde c}}}_i}} \right) = \,\bmod \,{{\mkern 1mu} _N}\left\{ {n_i^ *  - \frac{V}{2}, \cdots ,n_i^ *  + \frac{{V - 2}}{2}} \right\},
\end{equation}
where ${\text{Card}\left( \text{supp}\left( {{{\mathbf{\tilde{c}}}}_{i}} \right) \right)=V}$, and ${{\bmod _N}\left(  \cdot  \right)}$ is the modulo operation with respect to ${N}$, which guarantees that all indices in ${{\rm{supp}}\left( {{{{\bf{\tilde c}}}_i}} \right)}$ belong to ${\left\{ {1,2, \cdots ,N} \right\}}$. After the support of ${{{{{\bf{\tilde c}}}_i}}}$ has been detected, we can extract ${V}$ columns from ${{\bf{\bar W}}}$~(\ref{eq15}) according to ${{\rm{supp}}\left( {{{{\bf{\tilde c}}}_i}} \right)}$, and use the classical LS algorithm to estimate the nonzero elements of ${{{{{\bf{\tilde c}}}_i}}}$.

\begin{figure}[tp]
\begin{center}
\vspace*{0mm}\includegraphics[width=0.85\linewidth]{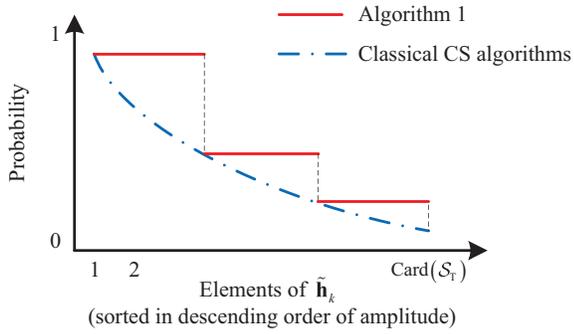}
\end{center}
\vspace*{-5mm}\caption{An illustration of the probability comparison.} \label{FIG3}
\end{figure}

Based on the discussion so far, the pseudo-code of the proposed SD-based channel estimation can be summarized in \textbf{Algorithm 1}\footnote{Note that the proposed SD-based channel estimation can be easily extended to the scenarios where users employ multiple antennas. In this case, the beamspace channel is a block sparse matrix instead of a sparse vector~\cite{brady2013beamspace}, and we can use \textbf{Algorithm 1}  for every column of this 2D sparse matrix.}, which can be explained as follows. During the ${i}$th iteration, we first detect the position ${{p_i}}$ of the strongest element of ${{{{{\bf{\tilde c}}}_i}}}$ in step 1. Then in step 2, utilizing the structural characteristics of beamspace channel as analyzed above, we can directly obtain ${{{\rm{supp}}\left( {{{{\bf{\tilde c}}}_i}} \right)}}$ according to~(\ref{eq24}). After that, the nonzero elements of ${{{{{\bf{\tilde c}}}_i}}}$ are estimated by LS algorithm in step 3, and the influence of this channel component  is removed in steps 4 and 5. Such procedure will be repeated (${i=i+1}$ in step 6) until the last channel component  is considered.
Note that for the proposed SD-based channel estimation, we do not directly estimate the beamspace channel as ${{\bf{\tilde h}}_k^{\rm{e}} = \sqrt {\frac{N}{L + 1}} \sum\limits_{i = 0}^L {{\bf{\tilde c}}_i^{\rm{e}}}}$. This is because that most of the elements with small power are regarded as zero, which will lead to error propagation in the influence removal, especially when ${i}$ is large. As a result, ${{\bf{\bar z}}_k^{\left( {i} \right)}}$ will be more and more inaccurate to estimate the nonzero elements in step 3. To this end, we only utilize ${{\bf{\bar z}}_k^{\left( {i} \right)}}$ to estimate the position ${{p_i}}$ in step 1, which can still guarantee a high recovery probability even if ${{\bf{\bar z}}_k^{\left( {i} \right)}}$ is inaccurate~\cite{tropp2007signal}. Then, after the iterative procedure, we can obtain the total support ${{{\cal S}_{\rm{T}}}}$ of ${{{\bf{\tilde h}}_k}}$  in step 7. Using ${{{\cal S}_{\rm{T}}}}$ and ${{{\bf{\bar z}}_k}}$, we can alleviate the impact of error propagation and estimate the beamspace channel more accurately in steps 8 and 9.

\begin{algorithm}[tp]
\caption{Proposed SD-based channel estimation.}
\KwIn{\\
\\\hspace*{+2mm} Measurement vector: ${{{\bf{\bar z}}_k}}$ in~(\ref{eq15});
\\\hspace*{+2mm} Combining matrix: ${{{\bf{\bar W}}}}$ in~(\ref{eq15});
\\\hspace*{+2mm} Total number of channel components: ${L + 1}$;
\\\hspace*{+2mm} Retained number of elements for each component: ${V}$.}
\textbf{Initialization}: ${{\bf{\tilde c}}_i^{\rm{e}} = {{\bf{0}}_{N \times 1}}}$ for ${0 \le i \le L}$, ${{\bf{\bar z}}_k^{\left( 0 \right)} = {{\bf{\bar z}}_k}}$.
\\\textbf{for} ${0 \le i \le L}$
 \\1. Detect the position of the strongest element in ${{{{{\bf{\tilde c}}}_i}}}$ as
 \\\hspace*{+4mm} ${{p_i} = \mathop {\arg \max }\limits_{1 \le n \le N} \left| {{{{\bf{\bar w}}}_n}{\bf{\bar z}}_k^{\left( i \right)}} \right|}$, ${{{{{\bf{\bar w}}}_n}}}$ is the ${n}$th row of ${{{\bf{\bar W}}}}$;
 \\2. Detect ${{\rm{supp}}\left( {{{{\bf{\tilde c}}}_i}} \right)}$ according to~(\ref{eq24});
 \\3. LS estimation of the nonzero elements of ${{{{{\bf{\tilde c}}}_i}}}$ as
 \\\hspace*{+4mm} ${{{\bf{f}}_i} = {\left( {{{{\bf{\bar W}}}_i}{\bf{\bar W}}_i^H} \right)^{ - 1}}{{\bf{\bar W}}_i}{\bf{\bar z}}_k^{\left( i \right)}}$, ${{{\bf{\bar W}}_i} = {\bf{\bar W}}{\left( {l,:} \right)_{l \in {\rm{supp}}\left( {{{{\bf{\tilde c}}}_i}} \right)}}}$;
 \\4. Form the estimated ${{\bf{\tilde c}}_i^{\rm{e}}}$ as ${{\bf{\tilde c}}_i^{\rm{e}}\left( {{\rm{supp}}\left( {{{{\bf{\tilde c}}}_i}} \right)} \right) = {{\bf{f}}_i}}$;
 \\5. Remove the influence of ${{{{{\bf{\tilde c}}}_i}}}$ as ${{\bf{\bar z}}_k^{\left( {i + 1} \right)}  = {\bf{\bar z}}_k^{\left( i \right)} - {{\bf{\bar W}}^H}{\bf{\tilde c}}_i^{\rm{e}}}$
 \\6. ${i=i+1}$;
\\\textbf{end for}
\\7. ${{{\mathcal{S}}_{\text{T}}}=\underset{0\le i\le L}{\mathop{\bigcup }}\,\text{supp}\left( {{{\mathbf{\tilde{c}}}}_{i}} \right)}$;
\\8. ${{{\bf{f}}_{\rm{T}}} = {\left( {{{{\bf{\bar W}}}_{\rm{T}}}{\bf{\bar W}}_{\rm{T}}^H} \right)^{ - 1}}{{\bf{\bar W}}_{\rm{T}}}{{\bf{\bar z}}_k}}$, ${{{\bf{\bar W}}_{\rm{T}}} = {\bf{\bar W}}{\left( {l,:} \right)_{l \in {{\cal S}_{\rm{T}}}}}}$;
\\9. ${{\bf{\tilde h}}_k^{\rm{e}} = {{\bf{0}}_{N \times 1}}}$, ${{\bf{\tilde h}}_k^{\rm{e}}\left( {{{\cal S}_{\rm{T}}}} \right) = {{\bf{f}}_{\rm{T}}}}$;
\\\KwOut{Estimated beamspace channel for user ${k}$: ${{\bf{\tilde h}}_k^{\rm{e}}}$.}
\end{algorithm}

The key difference between \textbf{Algorithm 1} and classical CS algorithms~\cite{tropp2007signal} is the step of support detection. In classical CS algorithms, all the positions of nonzero elements are estimated in an iterative procedure, which may be inaccurate, especially for the element whose power is not strong enough. By contrast, in our algorithm, we only estimate the position of the strongest element. Then, by utilizing the structural characteristics of mmWave beamspace channel, we can directly obtain the accurate support with higher probability as illustrated in Fig. 4. Moreover, we can also observe that the most complicated part of the proposed SD-based channel estimation is the LS algorithm, i.e., step 3 and step 8. Therefore, the computational complexity of SD-based channel estimation is ${{\cal O}\left( {L{V^2}M} \right){\rm{ + }}{\cal O}\left( {{\rm{Car}}{{\rm{d}}^2}\left( {{{\cal S}_{\rm{T}}}} \right)M} \right)}$ (${\text{Card}\left( {{\mathcal{S}}_{\text{T}}} \right)\le VL}$), which is comparable with that of LS algorithm, since ${L}$ and ${V}$ are usually small as  discussed above.

\section{Simulation Results}\label{S4}
In this section, we consider a typical mmWave massive MIMO system, where the BS equips a lens antenna array with ${N=256}$ antennas and ${{N_{{\rm{RF}}}}{\rm{ = 16}}}$ RF chains to simultaneously serve ${K=16}$ users. For the ${k}$th user, the spatial channel is generated as follows~\cite{sayeed2013beamspace,brady2013beamspace}: 1) one LoS component and ${L=2}$ NLoS components; 2) ${\beta _{k}^{\left( 0 \right)}\sim\mathcal{C}\mathcal{N}\left( 0,1 \right)}$, and ${\beta _{k}^{\left( i \right)}\sim\mathcal{C}\mathcal{N}\left( 0,{{10}^{-2}} \right)}$ for ${i = 1,2}$; 3) ${\psi _{k}^{\left( 0 \right)}}$ and ${\psi _{k}^{\left( i \right)}}$  follow the i.i.d. uniform distribution within ${\left[ -0.5,0.5 \right]}$.

Fig. 5 shows the normalized mean square error (NMSE) performance comparison between the proposed SD-based channel estimation and the conventional OMP-based channel estimation (i.e., using OMP to solve~(\ref{eq15})), where the total number of instants ${Q}$ for pilot transmission is ${Q = 96}$ (i.e., ${M = 6}$ blocks). For SD-based channel estimation, we retain ${V=8}$ strongest elements as analyzed above for each channel component, while for OMP, we assume that the sparsity level of the beamspace channel is equal to ${V\left( L\text{+}1 \right)=24}$. From Fig. 5, we can observe that SD-based channel estimation enjoys much better NMSE performance than OMP-based channel estimation, especially when the uplink SNR is low (e.g., less than 15 dB). Since low SNR is the typical case  in mmWave communications before beamforming~\cite{alkhateeb2014channel}, we can conclude that the proposed SD-based channel estimation is more attractive for mmWave massive MIMO systems.

\begin{figure}[tp]
\begin{center}
\vspace*{-3mm}\includegraphics[width=0.95\linewidth]{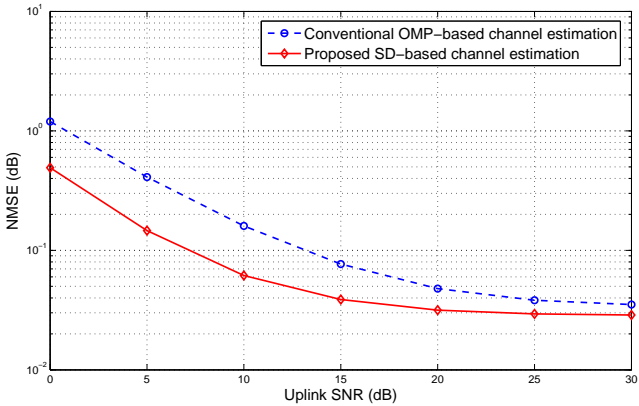}
\end{center}
\vspace*{-4mm}\caption{NMSE performance comparison.} \label{FIG3}
\end{figure}

\begin{figure}[tp]
\begin{center}
\vspace*{0mm}\includegraphics[width=0.95\linewidth]{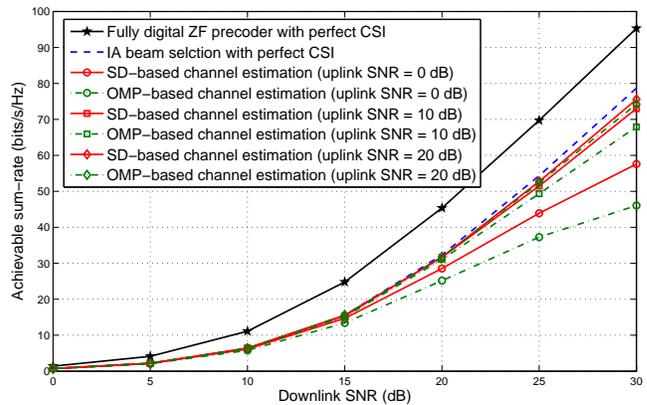}
\end{center}
\vspace*{-4mm}\caption{Sum-rate performance comparison.} \label{FIG3}
\end{figure}

Next, we evaluate the impact of different beamspace channel estimation schemes on beam selection. We adopt the interference-aware (IA) beam selection proposed in~\cite{gao16bs} as it can support the ${{N_{{\rm{RF}}}}=K}$ scenario, and the dimension-reduced digital precoder ${{{\bf{P}}_{\rm{r}}}}$ in~(\ref{eq5}) is selected as the zero-forcing (ZF) precoder. Fig. 6 provides the sum-rate performance of IA beam selection with different channels. We can observe that by utilizing the proposed SD-based channel estimation, IA beam selection can achieve better performance, especially when the uplink SNR is low. More importantly, when the uplink SNR is moderate (e.g., 10 dB), IA beam selection with SD-based channel estimation, which only requires 16 RF chains, can achieve the sum-rate performance not far away from the fully digital ZF precoder with 256 RF chains and perfect channel state information (CSI).

\vspace*{-1mm}
\section{Conclusions}\label{S5}
This paper investigates the beamspace channel estimation problem for mmWave massive MIMO systems with lens antenna array. Specifically, we first propose an adaptive selecting network with low cost to obtain the efficient measurements of beamspace channel. Then, we propose a SD-based channel estimation, where the key idea is to utilize the structural characteristics of mmWave beamspace channel to reliably detect the channel support. Analysis shows that the computational complexity of the proposed scheme is comparable with the classical LS algorithm. Simulation results verify that the proposed SD-based channel estimation can achieve much better NMSE performance than the conventional  OMP-based channel estimation, especially in the low SNR region.

\balance

\end{document}